\documentclass[11pt]{article}
\usepackage{a4}

\usepackage{amsfonts}
\usepackage{amsmath}
\usepackage{amssymb}

\newcommand{\bea}{\begin{eqnarray}}
\newcommand{\eea}{\end{eqnarray}}
\newcommand{\be}{\begin{equation}}
\newcommand{\ee}{\end{equation}}
\newcommand{\nn}{\nonumber}
\newcommand{\pa}{\partial}

\newcommand{\e}{\epsilon}

\newcommand{\cD}{{\cal D}}
\newcommand{\cH}{{\cal H}}

\newcommand{\hmu}{{\hat \mu}}
\newcommand{\hnu}{{\hat \nu}}
\newcommand{\hrho}{{\hat \rho}}
\newcommand{\hsigma}{{\hat \sigma}}

\newcommand{\hPhi}{\hat{\Phi}}
\newcommand{\hA}{\hat{A}}
\newcommand{\hB}{\hat{B}}

\newcommand{\hcH}{\hat{{\cal H}}}
\newcommand{\hLambda}{\hat{\Lambda}}
\newcommand{\hk}{\hat{\kappa}}
\newcommand{\hvphi}{\hat{\varphi}}
\newcommand{\hX}{\hat{X}}

\newcommand{\da}{\dot{a}}
\newcommand{\db}{\dot{b}}
\newcommand{\dc}{\dot{c}}
\newcommand{\dd}{\dot{d}}
\newcommand{\de}{\dot{e}}
\newcommand{\df}{\dot{f}}
\newcommand{\dg}{\dot{g}}

\newcommand{\dk}{\dot{k}}

\newcommand{\Ord}{{\cal O}}

\makeatletter

\@addtoreset{equation}{section}
\makeatother

\def\href#1#2{#2}

\begin{document}

\begin{titlepage}

\begin{center}

\hfill 
\vskip 1.1in

{\LARGE  Non-Linearly Extended Self-Dual Relations}\\[3mm] 
{\LARGE  From The Nambu-Bracket Description Of}\\[3mm]
{\LARGE  M5-Brane In A Constant $C$-Field Background}
\vskip 20mm
{\sc 
\large FURUUCHI \ Kazuyuki
}
\vskip 8mm
{\sl National Center for Theoretical Sciences\\
National Tsing-Hua University, Hsinchu 30013, Taiwan,
R.O.C.}\\
{\tt furuuchi@phys.cts.nthu.edu.tw}
\end{center}
\vspace{11pt}
\begin{abstract}
The derivation of the self-dual relations
for 
the two-form gauge field 
in the Nambu-bracket description
of M5-brane in a constant $C$-field background
initiated in Ref.\cite{Pasti:2009xc}
is completed by including contributions from
all the fields
in the M5-brane action.
The result is used to examine
Seiberg-Witten map of
the BPS conditions for the string solitons,
up to the first order in the expansion
by the parameter $g$ which characterizes
the strength of the interactions through the Nambu-bracket.
\end{abstract}

\end{titlepage}

\setcounter{footnote}{0}

\section{Introduction}

M-theory has been providing us important insights
into the non-perturbative aspects of string theory.
However, its microscopic definition
is still lacking.
M2-branes and M5-branes
are fundamental building blocks
of M-theory.
Considering the current status,
any new information on their properties
could be important for finding
more fundamental formulation of M-theory.

Recent few years have seen a rapid progress
in the description of M-theory branes:
A model for multiple M-theory membranes
with a symmetry
based on Lie 3-algebra
was proposed in
Refs.\cite{Bagger:2006sk,Bagger:2007jr,Gustavsson:2007vu}.
Starting from the BLG model,
an action for M5-brane was constructed
in Ho-Matsuo \cite{Ho:2008nn} 
and Ho-Imamura-Matsuo-Shiba \cite{Ho:2008ve}
by taking the Lie 3-algebra which is
defined through the Nambu-bracket \cite{Nambu:1973qe}.\footnote{%
Also see Ref.\cite{Gustavsson:2009qd} for a similar construction of
M5-brane action in a different background.}
This is in parallel with the construction of
a D$p$-brane in a constant $B$-field background
from infinitely many D($p-2$)-branes 
\cite{Connes:1997cr,Aoki:1999vr,Ishibashi:1998ni,Furuuchi:2002zy}.
The low energy effective action on the D$p$-brane
is given by Yang-Mills theory on non-commutative space.
In fact, soon after the 
discovery of the non-commutative description
of the D-brane worldvolume theory,
the uplift to M-theory, namely
M5-brane in a constant $C$-field background,
was also investigated by several groups.%
\footnote{A partial list includes
\cite{Gopakumar:2000ep,Bergshoeff:2000jn,Kawamoto:2000zt,Park:2000au,Gustavsson:2006ie,Ho:2007vk}.
Also see Ref.\cite{Chu:2009iv} for a study after the recent developments
of M-theory brane models.}
What was missing at the time
was the appropriate uplift of the non-commutative description
to that for the worldvolume theory of M5-brane, 
which now we have a candidate.

Interestingly, in the case of a D-brane in a constant $B$-field
background, 
there is also an
$S$-matrix equivalent description 
on a space with ordinary commutative coordinates.
The map between the non-commutative description
and the ordinary description is 
called Seiberg-Witten map \cite{Seiberg:1999vs}.\footnote{%
A similar map has appeared in the study
of lowest Landau level fermions \cite{Sakita:1993mc}.}
With this historical background in mind,
Ho et al. conjectured
that the Nambu-bracket description of
M5-brane 
is related to the previously found
ordinary description of M5-brane
\cite{Howe:1996yn,Howe:1997fb,Pasti:1997gx,Bandos:1997ui,%
Aganagic:1997zq,Bandos:1997gm,Bergshoeff:1998vx,Cederwall:1997gg}
in a constant $C$-field background
via a straightforward generalization of the 
Seiberg-Witten map.
The first non-trivial check of this conjecture 
was made in Ref.\cite{Furuuchi:2009zx}
for the BPS string-like soliton configurations on the M5-brane 
\cite{Howe:1997ue,Michishita:2000hu,Youm:2000kr},
which describe M2-branes ending on the M5-brane.

A peculiar feature of
the M5-brane action of Refs.\cite{Ho:2008nn,Ho:2008ve}
was that 
some components of the two-form gauge field
were absent.
In
Ref.\cite{Pasti:2009xc},
it was demonstrated 
how the missing components
of the two-form gauge field
as well as self-dual relations 
for the field strength of the two-form gauge field
can be obtained from
this M5-brane action.
The self-dual relations for
the two-form gauge field
are characteristic feature of
M5-brane, 
and how to describe 
the self-dual two-form gauge field
\cite{
Pasti:1995ii,Pasti:1995tn,%
Pasti:1996vs,Perry:1996mk,
Howe:1997vn}
was a central issue in the 
previous constructions of the M5-brane action
\cite{Pasti:1997gx,Bandos:1997ui,Aganagic:1997zq}.
It will be also important to clarify
how the self-dual relations are maintained 
in the Nambu-bracket description of M5-brane.

In this paper,
the derivation of the
self-dual relations
initiated in Ref.\cite{Pasti:2009xc} 
is completed
by including contributions from
all the fields in the Nambu-bracket
description of M5-brane.
The necessity of the inclusion of the scalar fields 
to the self-dual relations
can be understood
considering the fact that
the scalar fields are 
related to
the embedding coordinate fields
in the ordinary description of M5-brane via
the Seiberg-Witten map:
In the ordinary description of M5-brane,
the self-dual relations
involve the embedding coordinate fields 
through the induced metric on the M5-brane.
The obtained self-dual relations
in the Nambu-bracket description of M5-brane
are then used to examine the conjectured equivalence
between the Nambu-bracket description and the ordinary description
of M5-brane via the Seiberg-Witten map, 
in the case of 
the BPS conditions for 
the string solitons.
Since the string solitons involve
non-trivial configurations of a scalar field,
the inclusion of the contribution
of the scalar fields in the self-dual relations
is essential.

\section{The Nambu-bracket description of M5-brane
in a constant $C$-field background}

In this section I review 
the Nambu-bracket description of 
M5-brane in a constant $C$-field background
constructed in Ref.\cite{Ho:2008ve}
and fix my notation.
(Also see
Ref.\cite{Ho:2009zt}
for a concise review).

The action given in Ref.\cite{Ho:2008ve} describes
an M5-brane 
in the eleven-dimensional Minkowski space
whose worldvolume extends in one time and five spacial
directions.
The direction along the
worldvolume are parametrized by
coordinates
$x^{a}$ $(a=0,1,2)$ and
$y^{\da}$ $(\da = 3,4,5)$.
The metric on the worldvolume is specified by
the components
$\eta_{ab} = \mbox{diag} (-++)$, $\delta_{\da\db}$, 
and other components are zero.
There is a constant $C$-field background,
with only $C_{012}$ and $C_{345}$ components are non-zero.
Although both $C_{012}$ and $C_{345}$ components
are turned on,
the treatments of $012$ directions and $345$ directions
are quite asymmetric, which might be
one of the reasons why this action was not discovered
until recently.
The field content
of the Nambu-bracket description
of M5-brane is as follows:
The scalars $X^I$ $(I = 6,\cdots,10)$ 
describe embedding coordinates
transverse to the M5-brane worldvolume.
The six-dimensional chiral fermions 
can be conveniently parametrized
by a single eleven-dimensional Majorana spinor $\Psi$
satisfying
\bea
\Gamma \Psi = - \Psi,
\eea
where $\Gamma$ is given by
\bea
\Gamma = \Gamma^0\Gamma^1\Gamma^2\Gamma^3\Gamma^4\Gamma^5 .
\eea
The $\Gamma$-matrices are those for the eleven-dimensional space-time.
The salient feature of the M5-brane
worldvolume theory is the self-duality
of the two-form gauge field
$A$, which is the focus of this paper.
The components of the
self-dual two-form gauge field $A$ 
should be given by
$A_{ab}$, $A_{a\db}$, $A_{\da\db}$,
but the components $A_{ab}$ do not appear in the action.
They will appear from the equations of motion,
as will be described in the next section.

The M5-brane action 
is given as follows:
\bea
 \label{M5action}
S = \frac{T_{M5}}{g^2}
\left(
S_{B} + S_{CS} + S_{F}
\right) ,
\eea
where
\bea
S_{B}
&=&
\int d^3xd^3y
\Biggl[
-\frac{1}{2} \cD_{a} X^I \cD^{a} X^I
-\frac{1}{2} \cD_{\da} X^I \cD^{\da} X^I \nn \\
&&
-\frac{1}{4} \cH_{a\db\dc} \cH^{a\db\dc} 
-\frac{1}{12} \cH_{\da\db\dc} \cH^{\da\db\dc}
-\frac{g^4}{4}  \{X^{\da}, X^I, X^J \}\{X^{\da}, X^I, X^J \} \nn \\
&&
-\frac{g^4}{12}  \{X^I, X^J, X^K \}\{X^I, X^J, X^K \}
\Biggr] , \\
S_{CS}
&=&
-
\int d^3x d^3 y
\Biggl[
\frac{1}{2}
\e^{abc} B_a{}^{\da} \pa_b A_{c\da}
+
g \det B_a{}^{\da} 
\Biggr], \\
S_{F}
&=&
\int d^3x d^3y
\Biggl[
\frac{i}{2}
\bar{\Psi}
\Gamma^a \cD_a \Psi 
+
\frac{i}{2}
\bar{\Psi}
\Gamma^{\da} \cD_{\da} \Psi  \nn\\
&&
+
\frac{i g^2}{2}
\bar{\Psi}
\Gamma_{\da I} \{ X^{\da},X^I,\Psi \}
-
\frac{i g^2}{4}
\bar{\Psi}\Gamma_{IJ} \Gamma_{345} \{ X^I,X^J,\Psi \}
\Biggr] .
\eea
$\{\ast,\ast,\ast \}$ is the Nambu-bracket
on ${\mathbb R}^3$:
\bea
\{ f,g,h \}
=
\e^{\da\db\dc} 
\frac{\pa}{\pa y^{\da}} f \, \frac{\pa}{\pa y^{\db}} g  \, \frac{\pa}{\pa y^{\dc}} h ,
\eea
with $\e^{345}=1$.
The covariant derivatives in the action are given as
\bea
\cD_{a} \varphi &=& 
(\pa_a - g B_a{}^{\da} \pa_{\da} ) \varphi , \quad 
(\varphi = X^I, \Psi) , \label{cD} \\
\cD_{\da} \varphi &=& 
\frac{g^2}{2}
\e_{\da \db \dc}
\{X^{\db},X^{\dc},\varphi \} , \label{cDdot}
\eea
where
\bea
 \label{Xd}
X^{\da} = \frac{y^{\da}}{g} + A^{\da}, \quad A^{\da} = \frac{1}{2} \e^{\da\db\dc} A_{\db\dc} \, ,
\eea
and
\bea
B_a{}^{\da} = \e^{\da \db \dc} \pa_{\db} A_{a \dc} \, .
\eea
It follows that $\pa_{\dc} B_a{}^{\dc} = 0$.
When one derives the M5-brane action
from the BLG model, the components of the
two-form gauge field $A_{a \db}$
arise from the gauge field in the BLG model \cite{Ho:2008ve}.

The gauge transformation laws are given as
\bea
\delta_{\Lambda}
\varphi 
&=& g \kappa^{\dc} \pa_{\dc} \varphi, \quad (\varphi = X^I, \Psi), 
\label{gaugevphi}\\
\delta_{\Lambda} A_{\da\db} 
&=&
\pa_{\da} \Lambda_{\db} - \pa_{\db} \Lambda_{\da} + g \kappa^{\dc} \pa_{\dc} A_{\da\db} \, ,
\label{gaugeAdd}\\
\delta_{\Lambda} A_{a\db} 
&=&
\pa_{a} \Lambda_{\db} - \pa_{\db} \Lambda_{a} 
+ g \kappa^{\dc} \pa_{\dc} A_{a\db}
+ g (\pa_{\db} \kappa^{\dc} ) A_{a\dc} \, ,
\label{gaugeAd}
\eea
where
\bea
\kappa^{\da} = \e^{\da\db\dc} \pa_{\db} \Lambda_{\dc} \, .
\eea
It follows that $\pa_{\dc} \kappa^{\dc} = 0$.
Thus, the gauge transformation by the parameter
$\kappa$
generates volume-preserving diffeomorphisms,
and
$B_a{}^{\da}$ is the gauge field for the 
volume-preserving diffeomorphisms:
\bea
 \label{vd}
\delta_{\Lambda} y^{\da} = g \kappa^{\da} .
\eea
The transformation law of $B_a{}^{\da}$ under the 
volume-preserving diffeomorphisms
follows from eq.(\ref{gaugeAd}):
\bea
 \label{gaugeB}
\delta_{\Lambda} B_a{}^{\da}
=
\pa_a \kappa^{\da} 
+ 
g \kappa^{\db} \pa_{\db} B_a{}^{\da}
-
g B_{a}{}^{\db} \pa_{\db} \kappa^{\da} \, .
\eea
From eq.(\ref{gaugeB}) 
it follows that
the covariant derivatives 
(\ref{cD}) and (\ref{cDdot})
transform as scalars under the volume-preserving diffeomorphisms 
(\ref{vd}) \cite{Ho:2008ve,Pasti:2009xc}.
This 
allows one to construct gauge field strengths
which transform as scalars under the volume-preserving diffeomorphisms.

It is worth mentioning a subtle point here,
which was nicely explained in Ref.\cite{Pasti:2009xc}:
Although the fields $X^{\da}$ carry index ${\da}$,
they transform as scalars under the volume-preserving diffeomorphisms (\ref{vd}).
Indeed, in the derivation of the M5-brane action
from the BLG model in Ref.\cite{Ho:2008ve},
initially the target space indices of the scalar fields $X$ 
in the BLG model (with the Nambu-bracket as the Lie $3$-algebra)
have nothing to do
with the indices of the coordinates $y^{\da}$ on the 
Nambu-Poisson manifold (${\mathbb R}^3$ in current case).
What relates these different types of indices
is the background values of the scalar fields $X$:
\bea
 \label{bg}
X^{\da}_{bg} = \frac{y^{\da}}{g} , \quad \mbox{$\da = 3,4,5$} .
\eea
Since the identification of the indices in eq.(\ref{bg})
was made in a particular choice of coordinates,
it would appear different if one makes a volume-preserving
reparametrization of the coordinates.
However, one can keep the identification (\ref{bg}) intact instead, 
and absorb the induced change 
into the transformation of the 
``fluctuation" part $A^{\da}$ of the field $X^{\da}$ (\ref{Xd}).
This was how the gauge transformation law
(\ref{gaugeAdd}) arose from the scalar fields $X^{\da}$.\footnote{%
It may be useful to make a comment on 
a related but different topic for clarification.
One may try to compare the Nambu-bracket action of
M5-brane with the DBI-type action of M5-brane
which has {\em the worldvolume reparametrization invariance},
as discussed in the Discussions section 
(note that the BLG model with the Nambu-bracket 
only has the invariance under 
the volume-preserving diffeomorphisms 
in the dotted directions).
In this case,
$y^{\da}$ may be interpreted as (a part of)
the worldvolume coordinates of the M5-brane.
From the results in D-branes in a constant $B$-field background
\cite{Cornalba:1999hn,Cornalba:1999ah,Ishibashi:1999vi,Okuyama:1999ig,Jurco:2000fs},
it is expected that by choosing the so-called static {gauge}
$X^{\da} = y^{\da}/g$ 
for the worldvolume reparametrization 
of the DBI-type M5-brane action,
one obtains the ``commutative" description of M5-brane.
In this description, the fluctuation of the Nambu-Poisson structure
is parametrized by the ordinary gauge field in the DBI-type action.
On the other hand,
one can choose the gauge where the Nambu-Poisson structure is fixed.
The fluctuation part $A^{\da}$ of the scalar field 
cannot be eliminated in this gauge:
$X^{\da} = y^{\da}/g + A^{\da}$.
The volume-preserving part of the worldvolume reparametrization
remains as a residual symmetry,
and one obtains the Nambu-bracket description of M5-brane.
Please see the above mentioned papers for more detail
in the case of D-branes in a constant $B$-field background.
In the case of M5-brane, the detail has not been worked out at this moment.}

The field strengths for the two-form gauge field
which are made of the components
$A_{a\db}$ and $A_{\da\db}$ are given by
\bea
\cH_{a\db \dc} 
&=& \e_{\db \dc \dd} \cD_a X^{\dd} = F_{a\db \dc} - g B_a{}^{\dd} \pa_{\dd} A_{\db \dc} \, , 
\label{Hdd}\\
\cH_{\da \db \dc} 
&=& 
g^2 \e_{\da \db \dc} 
\left( \{ X^3,X^4,X^5 \} - \frac{1}{g} \right) \nn \\
&=&
F_{\da \db \dc} 
+ g \e_{\da \db \dc} 
\left(
(\pa_{\df} A^{\df})\pa_{\dg}A^{\dg} - (\pa_{\df} A^{\dg})\pa_{\dg}A^{\df}
\right)  + g^2 \e_{\da \db \dc} \{A^3,A^4,A^5\} , \label{Hddd}
\eea
where $F_{a \db \dc}$ and $F_{\da \db \dc}$
are components of the linear part of the field strength:
\bea
F_{a \db \dc} &=& \pa_a A_{\db \dc} - \pa_{\db} A_{a \dc} + \pa_{\dc} A_{a \db} \, , \\
F_{\da \db \dc} &=& \pa_{\da} A_{\db \dc} + \pa_{\db} A_{\dc \da} + \pa_{\dc} A_{\da \db} \, .
\eea
As mentioned above, the field stengths (\ref{Hdd}) 
and (\ref{Hddd}) transform as scalars under
the volume-preserving diffeomorphisms (\ref{vd}).

It is convenient to define a matrix $M_{\da}{}^{\db}$ following Ref.\cite{Pasti:2009xc}:
\bea
\label{M}
M_{\da}{}^{\db} = g \pa_{\da} X^{\db}.
\eea
The matrix $M_{\da}{}^{\db}$
transforms as a vector with respect to the lower index $\da$:
\bea
\delta_{\Lambda}
M_{\da}{}^{\db}
=
g \kappa^{\dc} \pa_{\dc} M_{\da}{}^{\db}
+
g(\pa_{\da}\kappa^{\dc}) M_{\dc}{}^{\db} .
\eea
Because of this property
the matrix $M_{\da}{}^{\db}$ and its inverse
can be used as a ``bridge"
which converts
scalar quantities to vector quantities, and vice versa.
In particular, the following identity holds:
\bea
 \label{bridge}
\cD_{\da} \varphi =
\det M M_{\da}^{-1}{}^{\db} \pa_{\db} \varphi.
\eea

The equations of motion of $X^I$ and $\Psi$
following from the action (\ref{M5action}) are 
\bea
0
&=&
\cD_a \cD^a X^I
+
\cD_{\da} \cD^{\da} X^I \nn \\
&&+
g^4 \{ X^{\da},X^J,\{ X^{\da}, X^J, X^I \} \}
+
\frac{g^4}{2} \{ X^J,X^K,\{ X^J, X^K, X^I \} \} \nn \\
&&+
\frac{i g^2}{2}
\{
\bar{\Psi} \Gamma_{\da I} , X^{\da},\Psi
\}
+
\frac{i g^2}{2}
\{
\bar{\Psi} \Gamma_{IJ}\Gamma_{345} , X^{J},\Psi
\} ,
\label{eqX} \\
0&=&
\Gamma^a \cD_a \Psi 
+
\Gamma^{\da} \cD_{\da} \Psi
+
g^2 \Gamma_{\da I} 
\{ X^{\da}, X^I, \Psi \} 
-
\frac{g^2}{2}
\Gamma_{IJ}\Gamma_{345} \{X^I,X^J,\Psi \}  .
 \label{eqPsi}
\eea
The equations of motion of gauge fields
$A_{a\db}$ and $A_{\da\db}$
and the Bianchi identity can be written as
\bea
\cD_a \cH^{a \db \dc} + \cD_{\da} \cH^{\da \db \dc} &=& g J^{\db \dc},
\label{eqA1} \\
\cD_a \tilde{\cH}^{a b \dc} + \cD_{\da} \cH^{\da b \dc} &=& g J^{b \dc} ,
\label{eqA2} \\
\cD_a \tilde{\cH}^{a b c} + \cD_{\da} \tilde{\cH}^{\da b c} &=& 0 ,
\label{Bianchi}
\eea
where $\cH^{\da b \dc} =  - \cH^{b \da \dc}$ and
\bea
J^{\da\db} = J^{\da\db}_B + J^{\da\db}_F,\quad
J^{a\db} =J^{a\db}_B + J^{a\db}_F,
\label{J}
\eea
\bea
J^{\da\db}_B
&=&
g\left(\{X^I,\cD^{\da}, X^{\db} \} - ({\da} \leftrightarrow {\db})\right)
-\frac{g^3}{2}
\e^{\da\db\dc} \{X^I,X^J,\{X^I,X^J,X^{\dc}\} \} ,
\label{JBdd}\\
J_F^{\da\db}
&=&\frac{ig}{2}
\left(
\{\bar{\Psi}\Gamma^{\da},X^{\db},\Psi \} - ({\da} \leftrightarrow {\db})
\right)
+\frac{i g^2}{2}
\e^{\da\db\dc} \{ \bar{\Psi} \Gamma_{\dc I}, X^I, \Psi \},\label{JFdd} \\
J^{a\db}_B
&=&
g \{X^I, \cD^{a} X^I, X^{\db} \} ,
\label{JBd}\\
J^{a\db}_F
&=&\frac{i g}{2}
\{ \bar{\Psi} \Gamma^a, \Psi, X^{\db}\} .\label{JFd}
\eea
The Hodge dual on the six-dimensional M5-brane worldvolume
is defined through the
totally anti-symmetric tensor
$\e_{\mu\nu\rho\lambda\sigma\delta}$
$(\mu,\nu = 0,1,\cdots,5)$:
\bea
\e_{abc\da\db\dc} = - \e_{\da\db\dc abc}
= \e_{a\db\dc bc\da} = \e_{abc}\e_{\da\db\dc} \, ,
\eea
with $\e_{012}= - \e^{012}= -1$.
A three-form $\cH$ is said to be
(linearly)
self-dual when it satisfies
\bea
 \label{LSD}
\tilde{\cH}_{\mu\nu\rho} = {\cH}_{\mu\nu\rho} \, ,
\eea
where
\bea
\tilde{\cH}_{\mu\nu\rho}
=
\frac{1}{6} \e_{\mu\nu\rho\lambda\sigma\delta} \cH^{\lambda\sigma\delta} .
\eea

\section{Non-linearly extended self-dual relations
from the Nambu-bracket description of M5-brane}

In Ref.\cite{Pasti:2009xc},
it was shown how 
the missing components of the 
two-form gauge field 
as well as 
(the non-linear extension of)
the self-dual relations
appear from the  
Nambu-bracket description of M5-brane action.
Their analysis was restricted to
the two-form gauge field part of the action.
In the following,
I will extend their results 
by including contributions from
all the fields in the M5-brane action.
The readers are recommended to 
go through the relevant part of Ref.\cite{Pasti:2009xc}.

Below I explain calculations
involving only bosonic fields in some detail.
Calculations involving 
fermions are bit lengthy but similar,
so I will just quote the final result 
in the appendix \ref{completeSDR}.

Following Ref.\cite{Pasti:2009xc},
I start from multiplying $M^{-1}_{\dc}{}^{\dd}$
to eq.(\ref{eqA2}),
where the matrix $M_{\da}{}^{\db}$ 
was defined in eq.(\ref{M}):
\bea
 \label{s11}
M^{-1}_{\dc}{}^{\dd}\cD_a \tilde{\cH}^{ab\dc}
+
M^{-1}_{\dc}{}^{\dd}\cD_{\da} \cH^{\da b\dc}
=
g M^{-1}_{\dc}{}^{\dd} J^{b\dc}_B .
\eea
Here, as mentioned above,
I consider the case
involving only bosonic fields,
so only $J^{b\dc}_B$ in $J^{b\dc}$ (\ref{J}) is taken into account.
In Ref.\cite{Pasti:2009xc}, 
it was shown that
the left hand side 
of eq.(\ref{s11}) can be written as a total derivative:
\bea
 \label{s12}
&&M^{-1}_{\dc}{}^{\dd}\cD_a \tilde{\cH}^{ab\dc}
+
M^{-1}_{\dc}{}^{\dd}\cD_{\da} \cH^{\da b\dc} \nn\\
&=&
\frac{1}{2} 
\e^{\da\db\dd}
\pa_{\da}
(
M_{\db}{}^{\df} \e_{\df\dg\dk} \cH^{b\dg\dk}
)
-
\e^{bac}\e^{\da\db\dd}
\pa_{\da}
\left(
\pa_{a} A_{c\db} + \frac{g}{2} \e_{\db \df\dg}B_{a}{}^{\df}B_{c}^{\dg} 
\right).
\eea
On the other hand, the right hand side of eq.(\ref{s11}) can also be 
written as a total derivative:
\bea
 \label{s13}
g M^{-1}_{\dc}{}^{\dd} J^{b\dc}_B
&=&
g^2 M^{-1}_{\dc}{}^{\dd} 
\e^{\de \df \dg}
(\pa_{\de} X^I )
(\pa_{\df} D^b X^I )
\pa_{\dg} X^{\dc} \nn \\
&=&
g
\e^{\de \df \dd}
(\pa_{\de} X^I )
\pa_{\df} D^b X^I 
=
g
\pa_{\df}
(
\e^{\de \df \dd}
(\pa_{\de} X^I) D^b X^I 
) \nn \\
&=&
- \e^{bac} \e^{\da \db \dd}
\pa_{\da}
\left(
\frac{g}{2}\e_{acd}
(\pa_{\db} X^I) D^d X^I 
\right) .
\eea
Notice the convention for the anti-symmetric tensor $\e^{abc}$
(see appendix \ref{conve}).
Eq.(\ref{s12}) and eq.(\ref{s13})
are total derivatives. 
By the Poincar\'{e} lemma
one obtains
\bea
 \label{SD1}
\cH^{a \db \dc}
=
\frac{1}{2} \e^{\db \dc \de}
\e^{abc}
M_{\de}^{-1 \dd}
\left(
F_{bc\dd} + g \e_{\dd \df \dg} B_{b}{}^{\df} B_{c}{}^{\dg}
- g \e_{bcd} (\pa_{\dd}X^I) \cD^d X^I 
\right) ,
\eea
where
\bea
F_{ab\dc} = \pa_a A_{b\dc} - \pa_b A_{a\dc} + \pa_{\dc} A_{ab}\, ,
\eea
and $A_{ab}(x,y)$ was introduced when integrating the
total derivative.
Eq.(\ref{SD1}) reduces to the linear self-dual relation
(\ref{LSD}) in the $g \rightarrow 0$ limit.
One may define 
a field strength as 
a scalar quantity with respect to the
area-preserving diffeomorphisms 
made only from the two-form gauge fields
which reduces to the linear field strength
in the $g \rightarrow 0$ limit:\footnote{%
In Ref.\cite{Pasti:2009xc} the authors 
imposed linear self-dual relations to define
$\cH_{ab\dc}$.
If one follows this reasoning here,
one needs to
define $\cH_{ab\dc}$ using scalar fields,
which seems little bit odd.
It would be more natural to define 
the field strength as in 
eq.(\ref{defHd}), and interpret
eq.(\ref{SD1}) as modified self-dual relations.
Note that in the ordinary description
of M5-brane, the self-dual relations
are also extended to the non-linear one
(the three-form field strength 
in the ordinary description can be related
to a self-dual three-form, which
might be closer to their identification).
In the case of $\cH_{abc}$ discussed later,
imposing the linear self-duality
does not uniquely fix the definition,
because the self-dual relation
includes a quadratic term in $\cH_{\da\db\dc}$,
see eq.(\ref{SD2}).
Anyway, it is not necessary to define the field strength
$\cH_{ab\dc}$ or $\cH_{abc}$ at this moment, 
and one can regard eq.(\ref{defHd}) as a shorthand notation
for the combination (\ref{SD1'}).}\label{footnoteH}
\bea
 \label{defHd}
\cH_{ab\dc}
=
M_{\dc}{}^{-1 \dd}
(
F_{ab\dd} + g \e_{\dd \de \df} B_{a}{}^{\de} B_{b}{}^{\df}
) .
\eea
Then, eq.(\ref{SD1}) takes the following form:
\bea
 \label{SD1'}
\cH^{a \db \dc}
=
\tilde{\cH}^{a \db \dc}
+ g \e^{\db\dc\de} M_{\de}{}^{-1 \dd} (\pa_{\dd}X^I) \cD^a X^I.
\eea
The appropriate gauge transformation law for
$A_{ab}$ which achieves the correct transformation property 
required by eq.(\ref{SD1}) is given as \cite{Pasti:2009xc}:
\bea
 \label{gaugeA}
\delta_{\Lambda} A_{ab}
&=&
\pa_a \Lambda_b - \pa_b \Lambda_a
+ 
g
(
\kappa^{\dc} \pa_{\dc} A_{ab}
+
 A_{a\dc} \pa_b \kappa^{\dc}
-
 A_{b\dc} \pa_a \kappa^{\dc} 
) \nn \\
&=&
\pa_a \Lambda_b - \pa_b \Lambda_a
+ 
g
(
\kappa^{\dc} \pa_{\dc} A_{ab}
-
(\pa_b A_{a\dc}) \kappa^{\dc}
+
(\pa_a  A_{b\dc}) \kappa^{\dc}
) \nn \\
&&+ \pa_b (gA_{a\dc} \kappa^{\dc}) - \pa_a (gA_{b\dc} \kappa^{\dc}) \nn\\
&=&
\pa_a \Lambda_b - \pa_b \Lambda_a
+ 
g F_{ab\dc} \kappa^{\dc} + \pa_b (gA_{a\dc} \kappa^{\dc}) - \pa_a (gA_{b\dc} \kappa^{\dc}).
\eea 
The last two terms in eq.(\ref{gaugeA})
can be absorbed in the redefinition of 
the gauge transformation parameter $\Lambda_a$.
The form in the last line in eq.(\ref{gaugeA})
is convenient for finding the Seiberg-Witten map 
for $A_{ab}$
which will be discussed in the next section.
From the gauge transformation law (\ref{gaugeA})
as well as (\ref{gaugevphi})--(\ref{gaugeAd}),
one can check that inside the parenthesis
of the right hand side of eq.(\ref{SD1})
transforms as a vector in the dotted directions.
Then, the multiplication of the matrix $M^{-1}$
converts this vector into a gauge scalar
(invariant under the gauge transformation generated by
$\Lambda_a$,
and scalar under the volume-preserving diffeomorphisms (\ref{vd})),
which is the same transformation property with 
the left hand side of eq.(\ref{SD1}), i.e. $\cH^{a\db\dc}$.

Next I look at eq.(\ref{eqA1}).
Multiplying
$M_{\da}{}^{\dd}\e_{\dd \db \dc}$ to eq.(\ref{eqA1}),
one obtains
\bea
 \label{s21}
M_{\da}{}^{\dd}\e_{\dd \db \dc}
\cD_a \cH^{a\db\dc}
+
2
M_{\da}{}^{\dd}
\cD_{\dd} \cH_{345}
=
g
M_{\da}{}^{\dd}\e_{\dd \db \dc}
J^{\db \dc}_B \, .
\eea
As noted before, I only
considered the bosonic fields above.
The second term 
in the left hand side of eq.(\ref{s21})
is equal to a total derivative:
\bea
2
M_{\da}{}^{\dd}
\cD_{\dd} \cH_{345}
=
\frac{1}{g}
\pa_{\da}
\left(
(\det M)^2 -1
\right),
\eea
where in the right hand side
the constant was introduced to ensure 
a smooth limit for $g\rightarrow 0$.
Here the identity (\ref{bridge}) has been used,
and notice that 
\bea
 \cH_{345} = \frac{1}{g} (\det M - 1 ) .
\eea
The first term in eq.(\ref{s21})
can be rewritten as
\bea
 \label{s23}
&& M_{\da}{}^{\dd}\e_{\dd \db \dc}
\cD_a \cH^{a\db\dc} \nn \\
&=&
\e^{abc}
M_{\da}{}^{\dd}
\cD_a 
(
\tilde{\cH}_{bc\dd}
- g \e_{bcd} M_{\dd}{}^{-1 \de}(\pa_{\de}X^I) \cD^d X^I
) \nn\\
&&+
2 M_{\da}{}^{\dd}
\cD_a
\left(
\cD^a X_{\dd} - \frac{1}{2} \e^{abc} 
(
\tilde{\cH}_{bc\dd} - g \e_{bcd} M_{\dd}{}^{-1 \de}(\pa_{\de}X^I) \cD^d X^I
)
\right) \nn\\
&=&
\e^{abc}
\cD_a
\left(
F_{bc\da}+g\e_{\dk\dg\da} B_{b}{}^{\dk} B_{c}{}^{\dg}
- g \e_{bcd} (\pa_{\da}X^I) \cD^d X^I
\right)
- 
2g (\cD_a \pa_{\da} X^{\dd}) D^a X_{\dd} \nn\\
&& 
+ 
2 
\cD_a
\left(
M_{\da}{}^{\dd}
 \left(
\cD^a X_{\dd} - \frac{1}{2} \e^{abc} 
(
\tilde{\cH}_{bc\dd}
- g \e_{bcd} M_{\dd}{}^{-1 \de}(\pa_{\de}X^I) \cD^d X^I
)
 \right)
\right). 
\eea
The last term in eq.(\ref{s23}) can be set to zero
by the self-dual relation (\ref{SD1}).
Compared with the result in Ref.\cite{Pasti:2009xc},
there is an extra term
\bea
&&
-\e^{abc} \cD_a
(g \e_{bcd} (\pa_{\da}X^I) \cD^d X^I ) \nn \\
&=&
-2g \cD_a ((\pa_{\da}X^I) \cD^a X^I )\nn \\
&=&
-g \pa_{\da} (D_a X^I D^a X^I) 
-2g^2 (\pa_{\da}B_{a}{}^{\dc}) (\pa_{\dc}X^I ) D^a X^I
-2g (\pa_{\da} X^I) D_a D^a X^I .
\label{s24}
\eea
The second term in the last line
in eq.(\ref{s24})
cancels a term
from the second term 
in the last but one line in
eq.(\ref{s23}),
the difference from Ref.\cite{Pasti:2009xc}
being the modification in 
the self-dual relation eq.(\ref{SD1}).
The last term of the last line in
eq.(\ref{s24})
can be rewritten
using the equation of motion for $X^I$ (\ref{eqX}),
with the fermions being set to zero:
\bea
&&-2g \pa_{\da} X^I D_a D^a X^I \nn\\
&=&
2g (\pa_{\da} X^I )
\Biggl[
\frac{g^4}{2}
\{X^{\dc},X^{\dd}, \{ X^{\dc},X^{\dd},X^I \} \}
+
g^4
\{X^{\dc},X^J, \{ X^{\dc},X^J,X^I \} \} \nn\\
&&
+
\frac{g^4}{2}
\{X^J,X^K, \{ X^J,X^K,X^I \} \}
\Biggr] .
 \label{s25}
\eea
Now I turn to the right hand side of eq.(\ref{s21}).
It can be rewritten as
\bea
&&g M_{\da}{}^{\dd}
\e_{\dd\db\dc} J^{\db \dc}_B \nn\\
&=&
g \pa_{\da} X^{\dd}
\e_{\dd\db\dc}
\Biggl[
g^2 
(
\{X^I,D^{\db}X^I,X^{\dc} \} - ({\db}\leftrightarrow {\dc})
)
-\frac{g^4}{2} \e^{\db\dc\de}
\{X^I,X^J,\{X^I,X^J,X_{\de} \}\}
\Biggr] \nn\\
&=&
-2g^5 \pa_{\da} X^{\dd}
\{X^{\dc},X^I,\{ X^{\dc},X^I,X^{\dd}\} \}
- 
g^5 \pa_{\da} X^{\dd}
\{X^I,X^J,\{ X^I,X^J,X^{\dd}\} \}.
\label{s26}
\eea
Eq.(\ref{s25}) and
eq.(\ref{s26})
are combined to give
\bea
&& -2g \pa_{\da} X^I D_a D^a X^I
- g M_{\da}{}^{\dd} \e_{\dd\db\dc} J^{\db \dc}_B \nn \\
&=&
-g^5
\pa_{\da}
\Biggl[
\frac{1}{2}
\{ X^{\dc},X^{\dd}, X^I \} \{ X^{\dc},X^{\dd}, X^I \}
+
\frac{1}{2}
\{ X^{\dc},X^I, X^J \} \{ X^{\dc},X^I, X^J \} \nn\\
&&+
\frac{1}{6}
\{ X^I, X^J, X^K \} \{ X^I, X^J, X^K \}
\Biggr]  \\
&&+ 2g^5 \{ X^{\dc},X^I, (\pa_{\da}X^{\dd})\{ X^{\dc},X^I, X^{\dd}\} \}
  +  g^5 \{ X^{\dc},X^{\dd}, (\pa_{\da}X^I)\{ X^{\dc},X^{\dd}, X^I\} \} \nn\\
&&+ 2g^5 \{ X^{\dc},X^J, (\pa_{\da}X^I)\{ X^{\dc},X^J, X^I\} \} 
  +  g^5 \{ X^I, X^J, (\pa_{\da}X^{\dd})\{ X^I,X^J, X^{\dd}\} \} , \nn
\label{s27}
\eea
where the derivation property of the
Nambu-bracket (see eq.(\ref{deri}) in the appendix) has been used.
The last two lines in eq.(\ref{s27})
can also be rewritten
as total derivatives
similar to the upper lines
using eq.(\ref{f1}) in the appendix.
Thus, 
I obtain 
the following non-linearly extended
self-dual relation:
\bea
 \label{SD2}
0&=&\frac{1}{3} \e^{abc} F_{abc} 
- g \e^{abc} B_{a}{}^{\db} F_{bc\db}
- 4g^2 \det B_a{}^{\db} -\frac{g}{2} \cH_{a\db\dc} \cH^{a\db\dc}
+ \frac{1}{g} \left( (\det M)^2 - 1 \right) \nn\\
&&
- g \cD_a X^I \cD^a X^I + g \cD_{\da} X^I \cD^{\da} X^I \nn \\
&&
+ \frac{g^5}{2} \{X^{\da},X^I,X^J \} \{X^{\da},X^I,X^J \}
- \frac{g^5}{6} \{X^I,X^J,X^K \} \{X^I,X^J,X^K \} .
\eea
Again, if one takes the $g=0$ limit, eq.(\ref{SD2}) reduces
to the linear self-dual relation (\ref{LSD}).
Thus eq.(\ref{SD1}) and eq.(\ref{SD2})
are regarded as the non-linear extension of the 
linear self-dual relations.
One may define 
another component of the
field strength with the similar reasonings
for the definition (\ref{defHd})
(see the footnote \ref{footnoteH}):
\bea
 \label{defH}
\cH_{abc} = F_{abc} +
\frac{g}{2}
\e_{abc} \e^{def}
B_{d}{}^{\db} F_{ef\db} 
+
2 g^2   \e_{abc}
\det B .
\eea
One can check that
$\cH_{abc}$ in eq.(\ref{defH})
transforms as a scalar
under the gauge transformations
(\ref{gaugeAd}) and (\ref{gaugeA}).

Calculations including the contributions from
fermions are similar.
The complete results including fermions
are included in the appendix \ref{completeSDR}.

\section{Seiberg-Witten map of BPS conditions
for the string solitons}

In this section,
I examine the Seiberg-Witten map of
the BPS conditions
for the string solitons on M5-brane
which was studied in Ref.\cite{Furuuchi:2009zx}.
The non-linearly extended self-dual relations
(\ref{SD1}) and (\ref{SD2}), 
which include all the bosonic fields in the M5-brane action,
are essential
since the string solitons 
involve non-trivial configuration of a scalar field.

Only in this section,
the fields in the Nambu-bracket description
are denoted with $\hat{\ }$ on them,
in order to distinguish them from
the corresponding fields in
the ordinary description
which are denoted without $\hat{\ }$.

Seiberg-Witten map is a solution to the condition:
``Gauge transformations in the Nambu description
is compatible with gauge transformations in the ordinary description":
\bea
 \label{SWc}
\hat{\delta}_{\hLambda} \hPhi (\Phi) = 
\hPhi (\Phi + \delta_\Lambda \Phi) - \hPhi (\Phi) ,
\eea
where $\hPhi$ ($\Phi$) collectively represents
fields in the Nambu-bracket (ordinary) description of M5-brane.
The Seiberg-Witten map 
for the fields $\hvphi$ $(\hvphi = \hX^I,\hat{\Psi})$, 
${\hA}^{\da}$, ${\hB}_a{}^{\da}$
and the gauge transformation parameter ${\hk}^{\da}$
were obtained in Ref.\cite{Ho:2008ve}:
\bea
\hvphi &=& \varphi + g A^{\da} \pa_{\da} \varphi 
+ \Ord (g^2)  , \quad (\hvphi = \hX^I,\hat{\Psi}),
\label{SWvphi}\\
{\hA}^{\da} &=& 
A^{\da} 
+ \frac{g}{2} A^{\db} \pa_{\db} A^{\da}
+ \frac{g}{2} A^{\da} \pa_{\db} A^{\db} 
+ \Ord (g^2) ,
\label{SWAd}\\
{\hB}_a{}^{\da} 
&=&
B_a{}^{\da}
+ g A^{\db} \pa_{\db} B_a{}^{\da}
-\frac{g}{2} A^{\db} \pa_a \pa_{\db} A^{\da}
+ \frac{g}{2} A^{\da} \pa_a \pa_{\db} A^{\db}  \nn \\ 
&&
+ g (\pa_{\db} A^{\db})  B_a{}^{\da} 
- g (\pa_{\db} A^{\da})  B_a{}^{\db}
- \frac{g}{2} (\pa_{\db} A^{\db}) \pa_a A^{\da} \nn \\
&&+ \frac{g}{2} (\pa_{\db} A^{\da}) \pa_a A^{\db} 
+ \Ord (g^2) ,
\label{SWB} \\
{\hk}^{\da} 
&=& \kappa^{\da} + \frac{g}{2} A^{\db} \pa_{\db} \kappa^{\da}
+ \frac{g}{2} (\pa_{\db} A^{\db}) \kappa^{\da} 
- \frac{g}{2} (\pa_{\db} A^{\da}) \kappa^{\db}
+ \Ord (g^2) .
\label{SWk}
\eea
On the other hand, from the gauge transformation law
(\ref{gaugeA}) one obtains the Seiberg-Witten map for $\hA_{ab}$:
\bea
 \label{SWA}
\hA_{ab} =
A_{ab} + g A^{\dc} F_{ab\dc} + \Ord (g^2).
\eea
Here, I have absorbed the last two terms
into the redefinition of $\Lambda_a$.
Notice that the gauge transformations 
generated by $\Lambda_a$
do not transform $A_{\da\db}$ nor $B_a{}^{\da}$.

I'd like to examine
the BPS conditions for the string solitons which were
studied in Ref.\cite{Furuuchi:2009zx} 
(see also Ref.\cite{Low:2009de}):\footnote{%
The BPS conditions
are crucial to justify our analysis here,
with the similar reason
explained in Ref.\cite{Hashimoto:2000mt}
in the case of BPS monopoles on a D$3$-brane 
in a constant $B$-field background.
We are planning to present the detail of the scaling arguments
to justify the use of the Nambu-bracket M5-brane action,
analogous to the zero-slope limit 
of the open string theory in a constant $B$-field background 
discussed in Ref.\cite{Seiberg:1999vs},
in our forthcoming paper.}
\bea
 \label{BPS}
{\cal D}_\hmu 
\hat{X}^6 + \eta
\frac{1}{6}
\e_{\hmu}{}^{\hnu \hrho \hsigma} 
\hat{{\cal H}}_{\hnu \hrho \hsigma} = 0 ,
\eea
and other fields set to zero,
where $\eta = \pm 1$ and $\hmu,\hnu = 2,\cdots,5$.
From the Seiberg-Witten map
(\ref{SWvphi})--(\ref{SWA}) as well as the definitions (\ref{Hdd}) and (\ref{Hddd}),
one obtains
\bea
\hcH_{\da\db\dc} 
&=&
\cH_{\da\db\dc} 
+ g 
(
A^{\dd}\pa_{\dd}\cH_{\da\db\dc} + (\pa_{\dd}A^{\dd}) \cH_{\da\db\dc}
)
+
\Ord (g^2) ,
\label{H345}\\
\hcH_{a\db\dc} 
&=&
\cH_{a\db\dc} 
+ g 
(
A^{\dd}\pa_{\dd}\cH_{a\db\dc} + (\pa_{\dd}A^{\dd}) \cH_{a\db\dc}
)
+
\Ord (g^2),
\label{Hadd} \\
\cD_{\da} \hvphi
&=&
\pa_{\da} \varphi 
+ 
g 
(
A^{\dc}\pa_{\dc} \pa_{\da} \varphi
+(\pa_{\dc}A^{\dc}) \pa_{\da} \varphi
)
+ \Ord (g^2) , 
\label{Dphi}\\
\cD_{a} \hvphi
&=&
\pa_{a} \varphi 
+ 
g 
(
A^{\dc}\pa_{\dc} \pa_{a} \varphi
+(\pa_{a} A^{\dc}-B_{a}{}^{\dc}) \pa_{\dc} \varphi
)
+ \Ord (g^2) .
\label{Ddphi}
\eea
Using these formulas, from the $\hmu=2$ case of eq.(\ref{BPS}):
\bea
{\cal D}_2 \hX
+ \eta
\hcH_{345} = 0 ,
\eea
where $\hX \equiv \hX^6$,
one obtains
\bea
 \label{F345}
F_{345} = - \eta ( \pa_2 X  + \eta g \pa_{\hmu} X \pa^{\hmu} X ) + \Ord (g^2),
\eea
where the BPS conditions at $\Ord (g^0)$ 
have been used to rewrite the $\Ord (g)$ term in eq.(\ref{F345}).
On the other hand, from the $\hmu= \da$ case of eq.(\ref{BPS}):
\bea
{\cal D}_{\da} \hX
- \eta
\frac{1}{2}
\e_{\da\db\dc} \hcH^{2\db\dc} \, ,
\eea
one obtains
\bea
 \label{F2}
F_{2\da\db}
=
\eta
\e_{\da\db\dc} \pa^{\dc} X +
\Ord (g^2) ,
\eea
where again the BPS conditions at $\Ord (g^0)$
have been used to obtain the $\Ord (g)$ term.
Eq.(\ref{F345}) and eq.(\ref{F2})
should be
compared with the results 
in the ordinary description
which were obtained in Ref.\cite{Michishita:2000hu}:
\bea
F_{012}+C_{012}
&=&
\eta_1
(\sin \theta + \cos \theta \pa_2 X), 
\label{MF012}\\
F_{01\da}
&=&
\eta_1 \cos \theta \pa_{\da} X,
\label{MF01} \\
F_{\da\db\dc} + C_{\da\db\dc}
&=&
\eta_2 \e_{\da\db\dc}
\left(
\pa_2 X +
\frac{\sin \theta (1+ (\pa_{\hmu} X \pa^{\hmu} X) )}{\cos \theta - \sin \theta \pa_2 X}
\right), 
\label{MF345}\\
F_{2\da\db}
&=&
-\eta_2 \e_{\da\db\dc} \pa^{\dc} X ,
\label{MF2}
\eea
where $C_{012}$ and $C_{\da\db\dc}$
are the components of the background $C$-field
in the ordinary description. 
Eq.(\ref{F345}) and eq.(\ref{F2})
match with 
eq.(\ref{MF345}) and eq.(\ref{MF2}) respectively 
in the case $\eta_1 \eta_2 = -1$,
with the identifications\footnote{%
This corrects sign errors in Ref.\cite{Furuuchi:2009zx}.
Notice that the background $C$-field is specified
by $\chi = \eta \theta$, so this identification
is independent of the choice of the BPS conditions (the choice of $\eta$ 
in eq.(\ref{BPS})), as it should be.}
\bea
 \label{pmap}
\chi \equiv \eta \theta =  g+ \Ord (g^2), \quad \eta=\eta_1.
\eea

To examine 
whether one can obtain eq.(\ref{MF012}) and eq.(\ref{MF01})
from the Nambu-bracket description
via the Seiberg-Witten map,
one needs to use
the non-linearly extended self-dual relations
eq.(\ref{SD1}) and eq.(\ref{SD2})
obtained in the previous section.
Substituting
eq.(\ref{H345})--(\ref{Ddphi}) 
into eq.(\ref{SD2}),
one obtains
\bea
 \label{F012F345}
F_{012} 
+
F_{345}
=
- g \pa_{\hmu} X \pa^{\hmu} X + \Ord (g^2) ,
\eea
where the BPS conditions and 
the self-dual relations
at $\Ord (g^0)$ has been used to rewrite
the $\Ord (g)$ term of eq.(\ref{F012F345}).
Again, eq.(\ref{F012F345}) matches with eq.(\ref{MF012})
and eq.(\ref{MF345})
with the identifications (\ref{pmap}), up to $\Ord (g)$.

To complete the check
at $\Ord (g)$,
one should check whether 
eq.(\ref{MF01})
can be obtained from the Nambu-bracket description.
In order to do this calculation,
one needs the expression
for the Seiberg-Witten map of $\hcH_{ab\dc}$,
which in turn requires the expression for
the Seiberg-Witten map 
of $\hA_{a\db}$ itself rather than the anti-symmetrized combination of its derivatives
$\hB_{a}{}^{\db} = \e^{\db\dc\dd} \pa_{\dc} \hA_{a\dd}$.
However, the Seiberg-Witten map 
of $\hA_{a\db}$ seems to require
a non-local expression \cite{Ho:2008ve}.
I leave this check to the future work.

\section{Discussions}

In this paper,
the derivation of  
the non-linearly extended self-dual relations
initiated in Ref.\cite{Pasti:2009xc}
was completed
by including contributions from all the
fields in the Nambu-bracket description
of the M5-brane action 
in a constant $C$-field background.
It is rather impressive 
that the procedure of Ref.\cite{Pasti:2009xc}
also works with the inclusion of all the fields in the action,
though it should work in order for the action
to describe M5-brane.
This result suggests
the existence of a formalism
with auxiliary fields and extra local symmetries
where the self-dual relations are more manifest,
which reduces to the currently discussed 
action upon gauge fixing.
The self-dual relations 
are characteristic feature of M5-brane,
and this result is of essential importance
when comparing
the Nambu-bracket description of M5-brane
to the ordinary description via the Seiberg-Witten map.

To compare the M5-brane action in the ordinary description
with the one in the Nambu-bracket description, 
it will be useful
to extend the new auxiliary field formalism
also introduced in Ref.\cite{Pasti:2009xc}
to the non-linear DBI-type action in the ordinary description,
so that it can be gauge fixed to the form
which is more convenient to compare with the M5-brane action 
in the Nambu-bracket description.
For this purpose, it will be useful to
understand the introduction the of auxiliary fields
in a systematic way.
An interesting work in this direction is made in Ref.\cite{Chen:2010jg}.

In these few years, 
several new formulations for
M-theory branes have been proposed, and
it is important to
examine
to what extent they can describe expected properties 
of M-theory branes.
Consistency with the reduction to type IIA string
is a necessary condition \cite{Ho:2008ve,Mukhi:2008ux,Gomis:2008uv,Benvenuti:2008bt,Ho:2008ei},
but it tends to hide the information of M-theory
which we are seeking for.
The comparison of 
the Nambu-bracket description of M5-brane
to the ordinary description
via the Seiberg-Witten map
is a direct check of the former as M-theory brane.
The relation to the
ordinary description,
which can be described in space-time covariant ways, 
will be important for 
the Lie 3-algebra to play
fundamental role
in the description of M-theory.
The importance of relating the 
BLG model to the covariant formulations was stressed in
Ref.\cite{Furuuchi:2008ki}, see also Ref.\cite{Furuuchi:2009ax}.
Another trial to relate 
the BLG model to the 
light-cone Hamiltonian of M5-brane
was made in Ref.\cite{Bandos:2008fr} without
the $C$-field background,
but only the Carrollian limit of 
the BLG model was obtained.
Another approach to M-theory branes
is the ABJM model of multiple membranes \cite{Aharony:2008ug}.
In Refs.\cite{Terashima:2008sy,Terashima:2009fy},
M5-brane solutions 
in the ABJM model 
were constructed.
To describe
M-theoretical or eleven-dimensional aspects
by these M5-brane solutions
one tends to encounter non-perturbative problems.\footnote{%
I thank Seiji Terashima for explaining this point.}
Such problems are certainly interesting, but also hard.
The approach from the Nambu-bracket description
of M5-brane has an advantage that
one can see the relation to 
the ordinary formulation at the classical level
through the Seiberg-Witten map.
On the other hand, 
the Seiberg-Witten map
has been solved
up to the first order in the expansion
by the parameter $g$
which characterizes the strength
of the interaction through the Nambu-bracket.
This is certainly not satisfactory,
and one would like to obtain all order expression.
Though interacting nature of the M2-brane
worldvolume theory makes the analysis complicated
compared with the open string worldsheet theory
on a D-brane
in a constant $B$-field background,
results in that case 
(see e.g. Refs.%
\cite{Cornalba:1999hn,Cornalba:1999ah,%
Ishibashi:1999vi,Okuyama:1999ig,Jurco:2000fs,%
Asakawa:1999cu,Liu:2000mja,Okawa:2001mv})
will give clues for   
how to obtain all order expression of the Seiberg-Witten map
in the case of M5-brane in a constant $C$-field background.
Another issue which calls for better understanding is that
in the case of a D-brane in a constant $B$-field background,
the product of fields is given by
Moyal product,
whereas the Nambu-bracket is an analogue of the 
Poisson-bracket, and the product is not defined.
To obtain the Moyal product description of D4-brane from
M5-brane via a compactification on a circle,
the Nambu-bracket should be deformed appropriately.
This issue is discussed in Ref.\cite{Chen:2010ny}.

\section*{Acknowledgments}
I am grateful to
Pei-Ming Ho for his question
which led me to investigate 
the self-dual relations in their M5-brane action, 
and Tomohisa Takimi
for useful discussions.
I am also thankful to them for showing their
drafts before publication.
I am also thankful to 
Takayuki Hirayama, Sheng-Yu Darren Shih 
and 
Dan Tomino for discussions
and to Wen-Yu Wen for reading the manuscript.
I would like to thank
Yukawa Institute for Theoretical Physics for the
hospitality and support during my stay for the
workshop ``Branes, Strings and Black Holes,"
where I could have useful discussions with
Yutaka Matsuo and Seiji Terashima
to whom I am also grateful.
It was a great pleasure to attend
the symposium
``Symmetry Breaking in Particle Physics"
in honor of Prof. Yoichiro Nambu
where I could have invaluable stimulations.
I would also like to thank
the organizers of the 
Taiwan String Theory Workshop 2010
for the invitation
which motivated me to look back our previous work
\cite{Furuuchi:2009zx} and extend the results.
I am also grateful to
Andreas Gustavsson and Shiraz Minwalla
for explaining their works at the workshop.
I would also like to thank Chuan-Tsung Chan for 
bringing some references to my attention.
This work is supported in part
by 
National Science Council of Taiwan
under grant No. NSC 97-2119-M-002-001.

\appendix

\section{Appendix}

\subsection{Convention for the totally anti-symmetric tensor $\e^{abc}$}\label{conve}

Convention (metric $\eta^{ab} = \mbox{diag}(-++)$):
\bea
\e^{012} = - \e_{012}  =1.
\eea
It follows that
\bea
 \frac{1}{2} \e_{abc} \e^{dbc} = - \delta_a^d, \quad
 \e_{abc} \e^{dec} = - \delta_a^d \delta_b^e + \delta_a^e \delta_b^d \, .
\eea
The determinant of a matrix $B_a{}^{\db}$ can be written as
\bea
 \det B_a{}^{\db} =
\frac{1}{6} \e^{abc}\e_{\da\db\dc} B_a{}^{\da} B_b{}^{\db} B_c{}^{\dc} \, .
\eea

\subsection{Complete form of the self-dual relations}\label{completeSDR}

\bea
 \label{SD1F}
\cH^{a \db \dc}
&=&
\frac{1}{2} \e^{\db \dc \de}
\e^{abc}
M_{\de}^{-1 \dd}
\left(
F_{bc\dd} + g \e_{\dd \df \dg} B_{b}{}^{\df} B_{c}{}^{\dg}
- g \e_{bcd} 
 \left(
(\pa_{\dd}X^I) \cD^d X^I
+ i (\pa_{\dd} \bar{\Psi}) \Gamma^d \Psi 
 \right)
\right) ,
\eea
\bea
 \label{SD2F}
0&=&\frac{1}{3} \e^{abc} F_{abc} 
- g \e^{abc} B_{a}{}^{\db} F_{bc\db}
- 4g^2 \det B_a{}^{\db} -\frac{g}{2} \cH_{a\db\dc} \cH^{a\db\dc}
+ \frac{1}{g} \left( (\det M)^2 - 1 \right) \nn\\
&&
- g  \cD_a X^I \cD^a X^I + g \cD_{\da} X^I \cD^{\da} X^I \nn \\
&&
+ \frac{g^5}{2} \{X^{\da},X^I,X^J \} \{X^{\da},X^I,X^J \}
- \frac{g^5}{6} \{X^I,X^J,X^K \} \{X^I,X^J,X^K \} \nn\\
&&
- i g
\bar{\Psi} \Gamma^{\da} \cD_{\da} \Psi
 - i g^3 \bar{\Psi}\Gamma_{\da I} \{X^{\da},X^I,\Psi \} 
+ \frac{i g^3}{2} \bar{\Psi}\Gamma_{IJ}\Gamma_{345} \{ X^I,X^J,\Psi \}.
\eea

\subsection{Some formulas for the Nambu-bracket}

Derivation property:
\bea
 \label{deri}
\{A,B,CD\} = 
\{A,B,C\} D + C\{A,B,D\} .
\eea
A frequently used formula:
\bea
 \label{f1}
  \{A \pa_{\da} B,C,D \} 
+ \{A \pa_{\da} C,D,B \}
+ \{A \pa_{\da} D,B,C \}
= 
\pa_{\da} 
\left(
A \{ B,C,D \}
\right) .
\eea
Here, it is assumed that
$A,B,C,D$ are bosonic quantities.
When some of them are fermionic,
one should assign sign factors appropriately.
Eq.(\ref{f1})
can be obtained from the identity
for the totally anti-symmetric tensor $\e^{\da\db\dc}$:
\bea
 \label{f2}
 \delta_{\da}^{\db} \e^{\df\dc\dd}
+\delta_{\da}^{\dc} \e^{\df\dd\db}
+\delta_{\da}^{\dd} \e^{\df\db\dc}
=
\delta_{\da}^{\df} \e^{\db\dc\dd} \, .
\eea

\bibliography{M5selfdual}
\bibliographystyle{utphys}

\end{document}